\documentclass[structabstract]{aa}  
\usepackage{graphicx}
\usepackage{natbib}
\bibpunct[]{(}{)}{;}{a}{}{,}
\usepackage{txfonts}

\begin{document}
\title{The substellar mass function in the central region of 
the open cluster Praesepe from deep LBT observations
\thanks{The LBT is an international collaboration among institutions in
Germany, Italy and the United States. LBT Corporation partners are: LBT
Beteiligungsgesellschaft, Germany, representing the Max-Planck Society, the
Astrophysical Institute Potsdam, and Heidelberg University; Istituto Nazionale
di Astrofisica, Italy; The University of Arizona on behalf of the Arizona
university system; The Ohio State University, and The Research Corporation, on
behalf of The University of Notre Dame, University of Minnesota and University
of Virginia.}}

\author{
W.~Wang \inst{1,2},
S.~Boudreault \inst{1,3,4},
B.~Goldman \inst{1},
Th.~Henning\inst{1},
J.~A.~Caballero \inst{5},
and C.~A.~L.~Bailer-Jones \inst{1},
}
\offprints{W. Wang, \email{wangw@nao.cas.cn}}

\institute{Max-Planck-Institut f\"ur Astronomie,
  K\"onigstuhl 17, D-69117 Heidelberg, Germany
  \and Key Laboratory of Optical Astronomy, National Astronomical Observatories,
  Chinese Academy of Sciences, Beijing 100012, China
  \and Visiting
  Astronomer at the Department of Physics and Astronomy, State
  University of New York, Stony Brook, NY 11794-3800, USA 
  \and Mullard
  Space Science Laboratory, University College London, Holmbury St
  Mary, Dorking, Surrey, RH5 6NT, United Kingdom 
  \and 
  Centro de Astrobiolog\'{\i}a (CSIC-INTA), Departamento de
  Astrof\'{\i}sica, PO Box 78, 28691 Villanueva de la Ca\~nada, Madrid,
  Spain
  }

\date{Received 17 August 2010 / Accepted 03 May 2011} 


\abstract
{Studies of the mass function (MF) of open clusters of different ages allow
  us to probe the efficiency with which brown dwarfs evaporate
  from clusters to populate the field. Surveys of older
  clusters (age\,$\gtrsim$\,100\,Myr) are not affected so severely by several problems
  encountered in young clusters, such as intra-cluster extinction and large
  uncertainties in brown dwarf models.}
{We present the results of a deep photometric survey to
  study the MF of the central region of the old open cluster Praesepe (age
  590$^{+150}_{-120}$\,Myr, distance 190$^{+6.0}_{-5.8}$\,pc),
  down to the substellar regime.}
{We performed an optical ($riz$ and $Y$-band) photometric survey of
Praesepe using the Large Binocular Telescope Camera covering an area of
0.59\,deg$^2$ in the cluster centre from $i\sim19.0$~mag ($\sim$100\,$M_{\rm
Jup}$) down to a 5$\sigma$ detection limit at $i\sim$25.6~mag
($\sim$40\,$M_{\rm Jup}$). The survey is approximately 95\% complete
at $i=23.8$~mag and $z=22.0$~mag ($\sim$55\,$M_{\rm Jup}$).
}
{We identify 59 cluster member candidates, of which 37 are substellar, by 
comparing with the predictions of a dusty atmosphere model. The MF of
those candidates rises from the substellar boundary until $\sim$67\,$M_{\rm
Jup}$ and then declines. This is quite different from the form inferred for
other open clusters older than 50\,Myr, but seems to be similar to those found
in very young open clusters, the MFs of which peak at $\sim$10\,$M_{\rm Jup}$. Either
Praesepe really does have a different MF from other clusters or they had
similar initial MFs but a different dynamical evolution.
Since most of the candidates are faint, we lack astrometric or
spectroscopic follow-ups to test their memberships. However, the
contaminations by field dwarfs, galaxies, or giants are found to have little
effect on the shape of MF and therefore the MF of `real' cluster
members should have similar characteristics. }
{}

\authorrunning{Wang, W. et al.}  \titlerunning{
  Mass function of Praesepe from deep LBT observations}
\maketitle

\section{Introduction}

The mass functions (MFs) of stellar and substellar populations have been
determined from optical and near-infrared surveys for several open clusters at
different ages, such as the Orion Nebula Cluster, $\sigma$~Orionis,
$\rho$~Ophiuchi, Taurus, IC~348, IC~2391, M35, the Pleiades, and the Hyades.
These MFs show clear heterogeneity (see Fig.~\ref{fig:all-mf}), which may be partially caused
by cluster evolution. 

Studies of relatively old open clusters (age\,$>$\,100\,Myr) are important for
two particular reasons: first, they allow us to study the
intrinsic evolution of basic properties of brown dwarfs (BDs), e.g., luminosity and effective
temperature, and to compare the evolution with structural and atmospheric
models; second, we may investigate how the BD and low-mass star populations as
a whole evolve, e.g., the efficiency with which BDs and low-mass stars
evaporate from clusters. These investigations have been carried out for the
Hyades (\citealt{bouvier2008} and references therein) and for Praesepe
(\citealt{boudreault2010} and references therein).

The Praesepe open cluster has been surveyed extensively in the past (cf.\
Table~\ref{tab01}), but only a few surveys have reached masses below the substellar
limit (and then only just). Several BD candidates were detected in those
surveys, some of which will be re-examined in the present work. The substellar
MF of Praesepe remains uncertain.

\cite{boudreault2010} observed a significant difference between the MFs of
Praesepe and Hyades. While they found that the Hyades MF has a maximum at
$\sim$0.6\,M$_\odot$ (\citealt{bouvier2008}), the MF of Praesepe continues to
rise from 0.8\,M$_\odot$ down to 0.1\,M$_\odot$. This is surprising, as both
clusters share similar physical properties (ages, mass, metallicity, and tidal
radii). Disagreement between the Praesepe and Hyades MFs could arise from
variations in the clusters' initial MFs, or from differences in their dynamical
evolution (\citealt{bastian2010}). Although different binary fractions could
cause the observed (system) MFs to differ, there is no clear evidence of any
variations in the  
binary fractions from measurements published in the literature
(\citealt{boudreault2010}).

In this paper, we present a survey of the very low-mass star and substellar
populations of Praesepe using the blue and red Large Binocular Cameras,
extending down to hitherto unexplored mass regimes ($\sim$40\,$M_{\rm Jup}$).
The main aims of our study are to search for new BDs and determine the MF of
the Praesepe for a large coverage of the substellar regime. 

The candidate selection procedure, and mass and temperature determination methods
employed in this study are similar to those adopted in \citet{boudreault2010}.
However, we probe a lower mass regime and use an evolutionary model
based on a dusty atmosphere instead of a combination of dust-free and dusty
models.

\begin{table*}

\caption{Summary of previous photometric surveys in the Praesepe around the
hydrogen-burning limit. The completeness limits correspond to a 10$\sigma$
detection while those of the present work and Boudreault et al. (2010)
correspond to 5$\sigma$ detections.}

\label{tab01}
\begin{center}
\begin{tabular}{lllll}
\hline
\hline
\noalign{\smallskip}
Authors 			& Telescope / instrument 		& Area  	& BD 		& Completeness limits \\
       				&           				&(deg$^2$)	& candidates	& (mag) \\  
\noalign{\smallskip}
\hline
\noalign{\smallskip}
Hambly et al. (1995)      	& COSMOS / POS \& UKSTU       		& 19       	& 0    		& $R_F \sim$20, $I_N \sim$19 \\ 
Pinfield et al. (1997)    	& INT / WFC        			& 1.0      	& $\sim$10$^{a}$  	& $R$=21.5, $I$=20.0, $Z$=21.5 \\
{Magazz\`u} et al. (1998)       & INT / WFC         			& 0.22     	& 1     	& $R$=22.2, $I$=21.2 \\ 
\citet{chappelle2005}           & INT / WFC                             & 2.6           & 4             & $I_c$=21.3, $Z$=20.5 \\ 
\citet{gonzalez-garcia2006}	& 3.5\,m CAHA / LAICA \& 5\,m Hale / LFI& 0.33		& 1  		& $i'$=23.8, $z'$=23.3 \\ 
Boudreault et al. (2010)  	& 3.5\,m CAHA / $\Omega$2k \& 2.2m La Silla / WFI 	& 3.1 		& 6 		& $I_c$=23.2, $J$=19.9, $Ks$=18.6 \\ 
This work$^{b}$                 & LBT / LBC 				& 0.59 		& 37 		& $r$=24.1, $i$=25.6, $z$=24.7, $Y$=20.3 \\ 
\noalign{\smallskip}
\hline
\noalign{\smallskip}
\end{tabular}
\end{center}
\begin{itemize}
\item[$^{a}$]  From Fig.~3 in Pinfield et al. (1997), about 10 of their 26 Praesepe member candidates
have masses below 72~$M_{\rm Jup}$.
\item[$^{b}$] Our work was complemented by the the $JK_{\it s}$ 3.5\,m CAHA / $\Omega$2k data from Boudreault et al. (2010).
\end{itemize}
\end{table*}

\section{\label{obs-data-calib} Observations and analysis}

\subsection{\label{obs} Observations}

The observations presented in this paper were carried out with the Large
Binocular Telescope (LBT) located on Mount Graham, Arizona
(\citealt{hill2006}), using the Large Binocular Cameras (LBCs, see
\citealt{Speziali2008}). The LBCs are two wide-field, high-throughput imaging
cameras, namely Blue (LBCB) and Red (LBCR), located at the prime focus stations
of the LBT. Each LBC has a wide field of view (23'$\times$23'), with
four CCD detectors of 2048$\times$4608 pixels each, providing images with a
sampling of 0.23\arcsec/pixel.

The optical design and detectors of the two cameras are optimized for different
wavelength ranges: one for ultraviolet--blue wavelengths (320--500\,nm,
including the Bessel $U$, $B$, $V$ and Sloan $g$ and $r$ bands), and one for
the red--infrared bands (500--1000\,nm, including the Sloan $i$, $z$ and
Fan $Y$ bands). In the full binocular configuration, both cameras are
available simultaneously, and both point in the same direction of the sky, thus
doubling the net efficiency of the LBT.

To accomplish the entire survey of the inner region of Praesepe, we carried out
three observing runs, in March 2008, December 2008, and February 2009.
Table~\ref{tab01a} summarizes the observations and Fig.~1 shows the areas
surveyed. The total area covered is 0.59~deg$^2$, about
1\,percent of the cluster region. The transmission curves of the
filters used in this survey is presented in Fig.~2, along with a synthetic
spectrum of a brown dwarf with $T_{\rm eff}$ = 2300\,K, log \textit{g} = 4.5 [CGS],
and solar metallicity (NextGen model).

\begin{figure}
\centering
  \includegraphics[width=8cm]{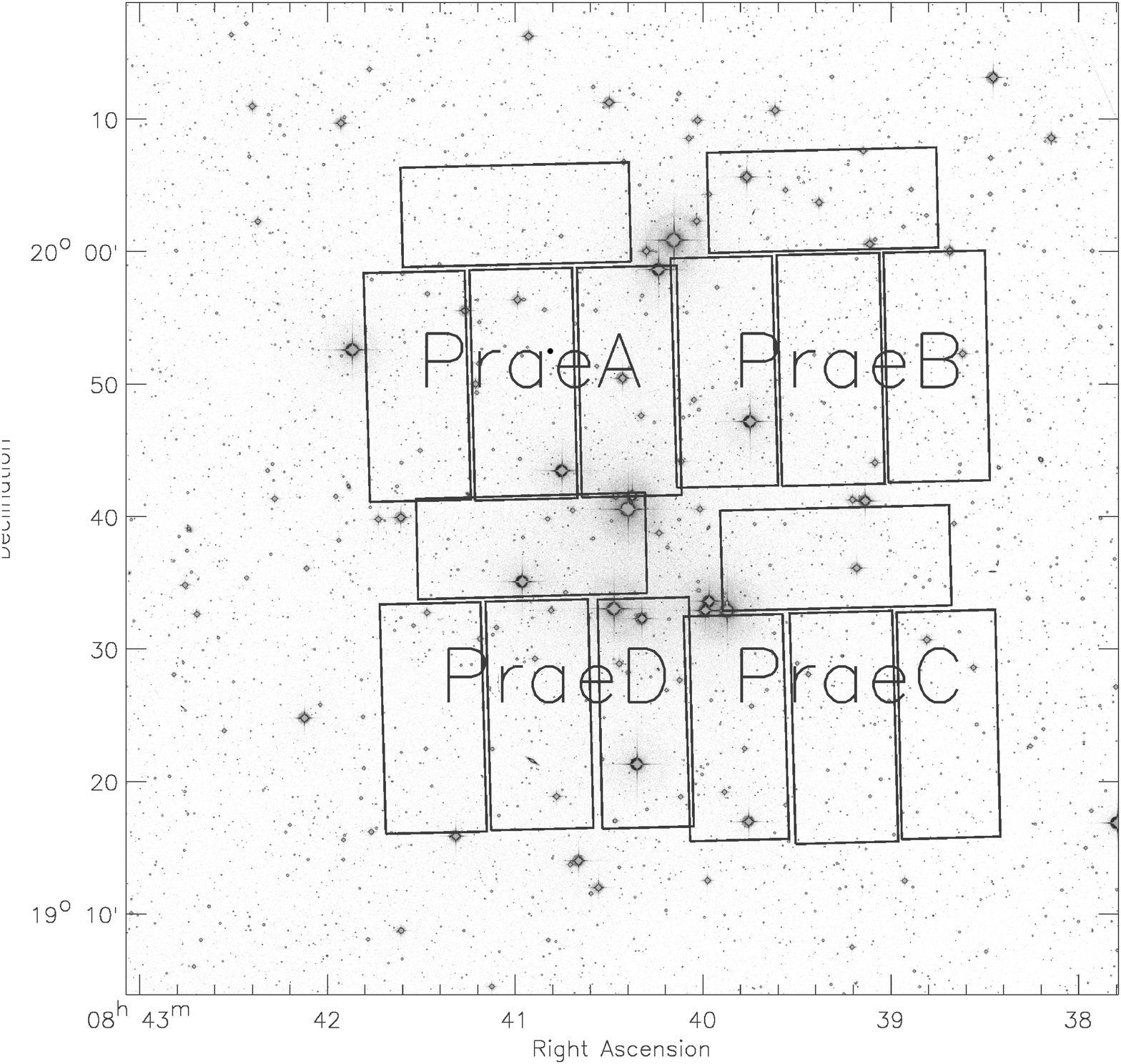}
  \caption{\label{f1} Spatial distribution of the LBT pointings}
\end{figure}

\begin{figure}
  \includegraphics[width=8cm]{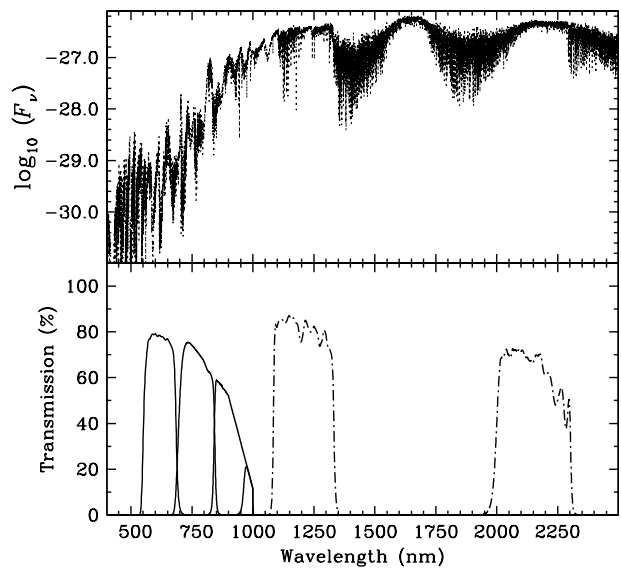}
  \caption{\label{fig:passband} Transmission curves of the filters used
    in our survey compared to a synthetic spectrum of a brown dwarf with
    $T_{\rm eff}$ = 2300\,K, log \textit{g} = 4.5, and solar metallicity
    (NextGen model). The transmission curves include the quantum
    efficiency of the detectors.
    }
\end{figure}

\begin{table*}
\caption{Journal of the LBT/LBC observations}
\label{tab01a}
\begin{center}
\begin{tabular}{l cc ccccc}
\hline
\hline
\noalign{\smallskip}
Field &RA          &DEC         &Date           & Filter&$t_{\it exp}$  &Seeing 	&$m_{5\sigma}$  \\
      &            &            & 	        &       &(min)  	&(arcsec)	&(mag)  	\\ 
\noalign{\smallskip}
\hline
\noalign{\smallskip}
PraeA &08:40:53.76 &+19:52:41.9 &2009-02-28     &$r$    &114		&1.0		&25.8		\\    
      &            &            &2009-02-28     &$i$    & 42		&1.0		&25.8		\\    
      &            &            &2009-02-28     &$z$    & 54		&1.0		&24.8		\\    
      &            &            &2009-02-28     &$Y$    & 18		&1.0		&21.3		\\    
\noalign{\smallskip}
PraeB &08:39:14.23 &+19:52:41.9 &2009-02-28     &$r^a$  & 54		&1.0		&25.1		\\
      &            &            &2008-03-06     &$i$    & 42		&2.2		&25.7		\\
      &            &            &2009-02-28     &$z$    & 54		&1.0		&25.2		\\
      &            &            &2008-03-04     &$Y$    & 48		&1.4		&22.2		\\
\noalign{\smallskip}
PraeC &08:39:14.36 &+19:27:18.0 &2008-03-07     &$r$    & 90		&1.9		&25.6		\\
      &            &            &2008-03-07     &$i$    & 30		&2.0		&25.8		\\
      &            &            &2008-03-07     &$z$    & 30		&1.0		&25.1		\\
      &            &            &2009-02-28     &$Y$    & 18		&1.0		&21.2		\\
\noalign{\smallskip}
PraeD &08:40:53.63 &+19:27:18.0 &2008-12-29     &$r$    & 99		&1.0		&25.8		\\
      &            &            &2008-12-29     &$i$    & 42		&1.0		&25.8		\\
      &            &            &2008-12-29     &$z$    & 39		&1.0		&24.8		\\
      &            &            &2008-12-29     &$Y$    & 18		&1.0		&21.9		\\
\noalign{\smallskip}
\hline
\noalign{\smallskip}
\end{tabular}
\begin{itemize}
\item[$^{a}$] We also performed a shallower, 84-min pointing in $r$ on 2008-12-30.
\end{itemize}
\end{center}
\end{table*}

\subsection{\label{data} Reduction and astrometry}

The standard data reduction steps for the LBT data were performed using the IDL
astronomy package and IRAF. The bias subtraction was executed on a nightly
basis and for each CCD chip. To correct for pixel-to-pixel variations and global
illumination, master flat frames were created for the nights using twilight
exposures. For nights when no appropriate sky flat exposures were
available, we used a master sky flat in the adjacent night. The individual
images of a given field were registered and median combined, resulting in a
combined science frame for each CCD, field and filter. To
detect faint sources, we subtracted the strong background introduced by
very bright stars$\footnote{The area fraction affected by bright stars 
is less than 3\% for most of CCD images, and is about 6\% in 
the worst case.}$. We then used the IRAF task {\tt daofind} to detect sources in the
``clean" frames. The sources were extracted from the original science frames and
instrumental magnitudes calculated using both aperture and point-spread function
photometry with the IRAF tasks {\tt phot} and {\tt allstar} respectively. 

At this stage, weak fringes were still visible for several $Y$-band images. The
method described by \citet{Bailer-Jones2001} for removing fringes does not
apply well in the present case, since the images are seriously affected by
bright stars and no clean fringe images could be created. However, as the
$Y$-band images are $\sim$1.5 dex shallower than expected, we decided
not to use them in this study (although we still quote some statistics
of the photometry below).

An astrometric solution was achieved using the Sloan Digital Sky Survey (SDSS
York et al. 2000) catalogue as a reference. The root mean square (rms)
accuracy of our astrometric solution is 0.10-0.15\,arcsec. As with 
other reduction procedures, astrometry was also performed separately for each
CCD, to ensure that the solutions were as robust as possible.

\subsection{\label{calib} Photometric calibration}

To correct for Earth atmospheric absorption of the photometry, we calibrated
the inferred data using the $r$, $i$, and $z$ band values of Sloan Digital Sky
Survey (SDSS) objects that were observed in the science fields. Zero point
offsets were determined from the difference between the SDSS magnitudes and our
instrumental magnitudes. Since these were obtained with objects in the same
field of view for each science frame, we did not perform a colour or airmass
correction when reducing our $riz$ photometry. The error introduced by this
approximation is less than 0.05~mag.

\begin{figure*}
\centering
\includegraphics[width=14cm]{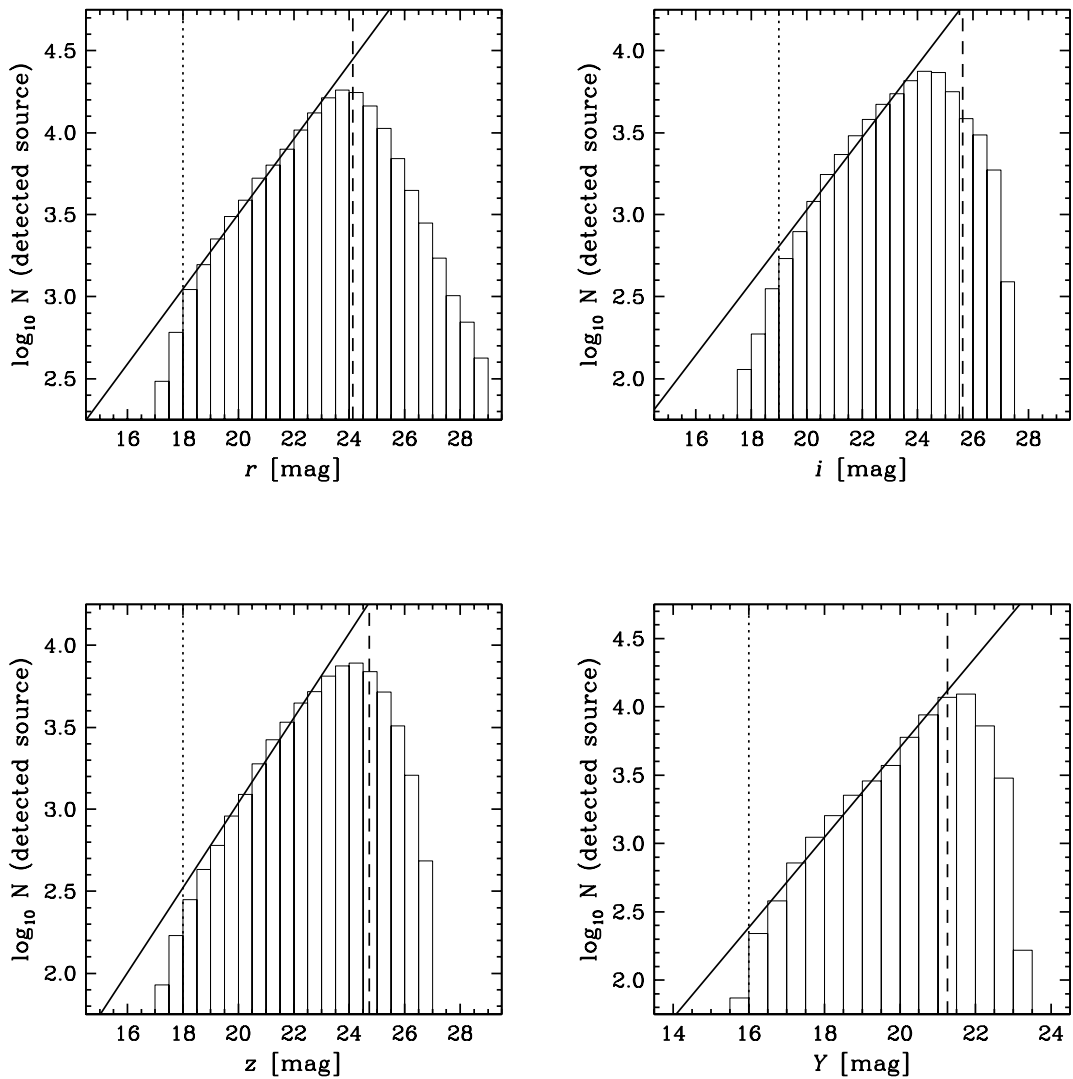}
\caption{\label{f6} Estimation of the completeness limit for our survey
using the $iz$ bands. The solid lines are the best linear
fits before the turn offs. The vertical dashed lines are the 5$\sigma$
detection limit and the vertical dotted lines are the saturation
magnitudes.}
\end{figure*}

To calibrate our $Y$ band photometry, we used our LBT $i$ and $z$
photometry and the $Y$ band photometry from the United Kingdom Infrared
Telescope Infrared Deep Sky Survey (UKIDSS, Lawrence et al. 2007). We found
that the differences between $Y$ band LBT magnitudes and $Y$ band UKIDSS
magnitudes have a linear dependence on the $i-z$ colours, which can be described
by the equation 
\begin{equation}
  Y_{UKIDSS}-Y_{LBT,raw}=a_0+a_1\times(i-z),
  \label{calib-eq1}
\end{equation}

On the basis of about 800 common objects between UKIDSS and our measurements, the
a$_{0,1}$ coefficients were determined and the instrumental $Y$ magnitudes were then
transferred into the UKIDSS $Y$ photometry system. For the same reasons as for our
$riz$ data, we did not (need to) perform a colour or airmass correction for our
$Y$-band photometry.

The 5$\sigma$ detection limits of our survey are $\sim$ 25.6\,mag and 24.7\,mag
for the $i$ and $z$ bands, respectively. However, we do not expect all targets
brighter than these limits to be able to be detected. We estimate the survey completeness
by comparing the number of objects detected to the number predicted assuming a
uniform three-dimensional spatial distribution of stars. As shown in
Fig.~\ref{f6}, the number of detected sources in each band deviates from a
log-normal relationship at bright and faint limits. From this, we estimate the
completeness to our 5$\sigma$ detection limit as 67.0\% at $i=25.6$ and
70.3\% at $z=24.7$ respectively, which corresponds to $\sim$40\,$M_{\rm Jup}$
assuming a dusty atmosphere. A similar estimation yields a completeness of
82.0\% at $r=24.1$\,mag and 82.6\% at $Y=20.3$\,mag, respectively. Our survey is
approximately 95\% complete at $i=23.8$~mag and $z=22.0$~mag. 

The relatively low completeness of our survey is possibly caused by the saturation
of bright stars. For stars lying near to saturated stars, the photometric
uncertainties are relatively large. A significant fraction of detected stars
is then excluded because of their large photometric errors, which lowers the
completeness. The total area seriously affected by bright stars in $i$ band is
3--6 percent.

\subsection{\label{selection} Candidate selection procedure}

The candidate selection introduced by \cite{boudreault2010} was
adopted in the present work. Candidates were first selected based on the
colour--magnitude diagram (CMD) using $iz$ bands from our LBC
observations\footnote{Our $rY$ bands observations do not reach a similar
stellar mass, hence are not used here.} A second selection was performed
using a colour--colour diagram. While our $rY$ bands observations are not deep
enough for our present investigations, data from the near-IR photometric survey by
\cite{boudreault2010} -- which fully covers our survey area, with a 5$\sigma$
detection limit at $\sim$55\,$M_{\rm Jup}$ in $J$ and $K_{\it s}$ bands -- was
used instead for the second selection. In the third and final selection, we
used the known distance to Praesepe to reject objects based on the discrepancy
between their observed magnitude in $J$ and the magnitude predicted from the
isochrones and our estimation of $T_{\rm eff}$. To be considered as a cluster
member, an object had to satisfy all three of these criteria.

We use the evolutionary tracks of \cite{chabrier2000} and the atmosphere
models from \cite{allard01} -- assuming a dusty atmosphere (the AMES-Dusty
model) -- to compute an isochrone for Praesepe using an age of
590$^{+150}_{-120}$\,Myr (\citealt{fossati2008}), a distance of
190$^{+6.0}_{-5.8}$\,pc (\citealt{Leeuwen2009}), and a solar metallicity
([Fe/H]\,=\,0.038$\pm$0.039, \citealt{friel1992}). We neglect the reddening
[$E(B-V)$\,=\,0.027$\pm$0.004\,mag, \citealt{taylor2006}]. The transmission
curves we used for the filters for these calculations are plotted in
Fig.~\ref{fig:passband}. The effective temperature varies from 500~K to 3900~K
in steps of 100~K, while the gravity ranges from 4.0~dex to 6.0~dex in steps of
0.5~dex. 

\subsubsection{\label{selection-1st} Colour-magnitude diagram}

\begin{figure}
  \centering
  \includegraphics[width=8cm]{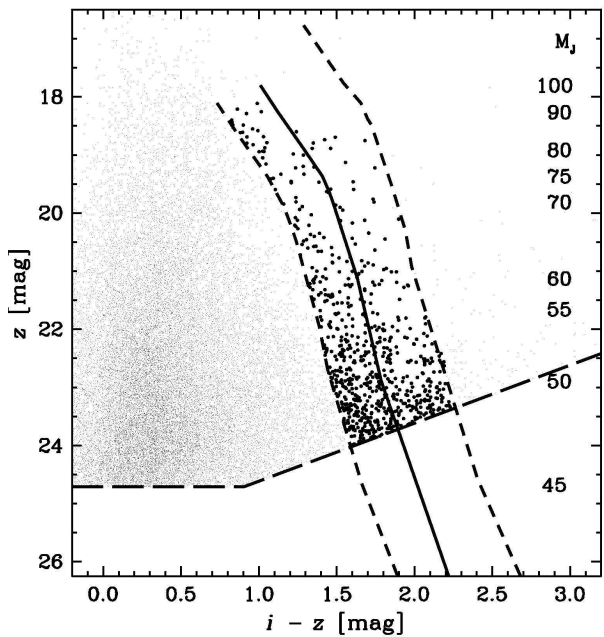}
  \caption{\label{fig:cmd-iz} CMD with $i$ and $z$ bands used in the
    first selection procedure. As solid lines we show the isochrone computed
    from an evolutionary model with a dusty atmosphere (AMES-Dusty).
    The numbers indicate the masses (in $M_{\rm Jup}$) on the model
    sequence for various $z$ magnitudes. The dashed lines delimit our
    selection band.}
\end{figure}

Candidates were first selected from our CMD by keeping all objects
that are no more than 0.28\,mag redder or bluer than the isochrones
in all CMDs. This number accommodates errors in the magnitudes and
uncertainties in the model isochrones. We also include
the errors from the age estimate and distance to Praesepe. We
additionally include objects brighter than the isochrones by
0.753\,mag in order to include unresolved binaries. In
Figure~\ref{fig:cmd-iz}, we show the CMD where candidates were selected
based on their values of $z$ versus $i$--$z$. 

From a total of 44\,209 objects above the 5$\sigma$ detection limit in $i$ and
$z$ filters and below our saturation limit, 709 objects are retained as
candidate cluster members (98.4\% are rejected). Of these, 160 were detected in
the NIR observations of \cite{boudreault2010}. These objects are used in
the selection process described below.

\subsubsection{\label{selection-2nd} Colour--colour diagram}

From this step, The candidate selection is based on both the optical $iz$ data
and the NIR $JK_{\it s}$ data. A candidate must be detected in every
band.

\begin{figure}
  \centering
  \includegraphics[width=8cm]{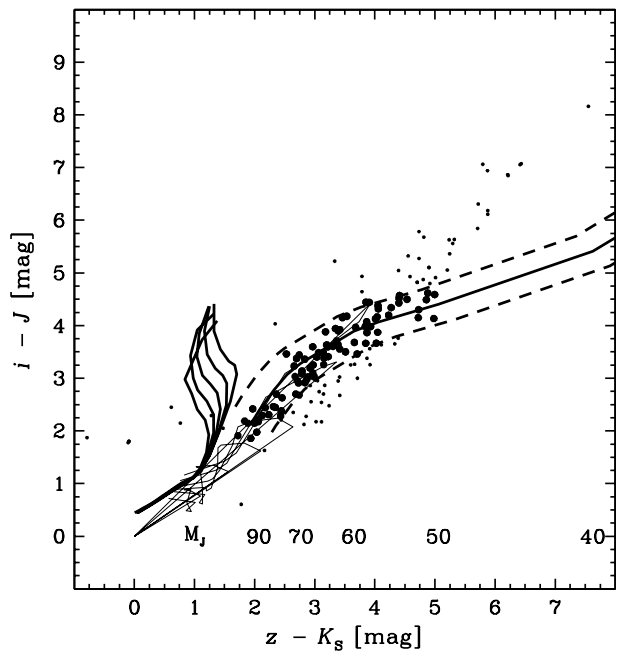}
  \caption{\label{fig:ccd-izjk} Colour-colour diagram used in the
    second selection step. The solid line is the isochrone computed
    from an evolutionary model with a dusty atmosphere (the AMES-Dusty model;
    the masses in M$_{\odot}$ for each $z-K_{\it s}$ colour
    are shifted for clarity). The dashed lines show our selection area. 
    We also show the theoretical colours of six galaxies as
    thin lines and the theoretical colours of red giants as
    thick lines. The six galaxies are two starbursts, one Sab,
    one Sbc, and two ellipticals of 5.5 and 15\,Gyr, with redshifts
    from $z$=0 to $z$=2 in steps of 0.25 (evolution not considered).
    We assume that all red giants have a mass of 5\,M$_{\odot}$, 0.5
    $<$ log \textit{g} $<$ 2.5 and 2000\,K $<$ $T_{\rm eff}$ $<$
    6000\,K.}
\end{figure}

The second stage of the candidate selection involves retaining only those
objects that lie within 0.28\,mag of the isochrone in the colour--colour
diagram.  This value accommodates the photometric errors, uncertainties in the
model isochrones, and the uncertainty in the age estimation of Praesepe. The
colour--colour diagram with the selection limits is shown in Figure
\ref{fig:ccd-izjk}, where we also plot the theoretical colours of red giants
\citep[using the atmosphere models of][]{hauschildt1999b} and the theoretical
colours of six galaxies with redshifts from 0 to 2 (Meisenheimer et al.\ 2011).
Neither the red giants nor the galaxies are expected to be a significant source
of contamination; most of the low redshift galaxies that were not automatically
discarded during PSF photometry with full-width-half-maxima (FWHMs) broader
than the average stellar FWHM by $\sim$30\%, were rejected by means of the
visual inspection of individual cluster member candidates after selection
procedures. Of the 160 objects selected in the first step, 88 are kept here.

\subsubsection{\label{selection-3rd} Observed magnitude vs.\ predicted magnitude}

As indicated in Section \ref{get-mass}, our determinations of $T_{\rm
eff}$ and mass are based on the spectral energy distribution of each object, so
are independent of the assumed distance. The membership status of an object can
therefore be assessed by comparing its observed magnitude in a band with its
magnitude predicted from its $T_{\rm eff}$ and Praesepe's isochrone (which
assumes a distance). The predicted magnitude of a background contaminant would
be lower (brighter) than its observed magnitude and higher (fainter) for a
foreground contaminant. To avoid removing unresolved binaries that are
real members of the cluster, we keep all objects with a computed magnitude of
up to 0.753\,mag brighter than the observed magnitude. We also take into
account photometric errors and uncertainties in the age and distance of
Praesepe. This selection procedure is illustrated in Figure \ref{fig:mj_vs_mj}.

Of the 88 objects selected through CMDs and colour-colour diagrams in the first
two steps, 74 are retained here. After this step, we perform a direct visual
inspection of the images to reject resolved galaxies and spurious
detections.  This inspection removes 15 objects from the photometric
selection, of which seven are possibly galaxies and eight are false detections.

The remaining objects constitute our final cluster member candidates, shown
as large dots in Fig.~\ref{fig:cmd-iz_izJK}. Those selected using only the $iz$
photometry amount to 709, and are presented in Fig.~\ref{fig:cmd-iz_izJK} as
small dots. We note that employing NIR $JK_{\it s}$ data helps us to remove a
significant fraction of contaminations. At each mass bin, we calculate the
number of stars removed as a result of including $JK_{\it s}$ data in the
selection, and use this to estimate the number of stars in the final mass
bin where $JK_{\it s}$ data are unavailable.

{\citet{Schmidt2010} and \citet{West2011} investigated the colors of L
and M dwarfs, respectively, for every spectral types using the SDSS and Two
Micron All Sky Survey (2MASS) catalogues. The observed colour
ranges are consistent with our computed colour ranges for M and early L
dwarfs. However, the observed $i-J$ color range for individual L dwarfs as
shown in Fig.2\, in \citet{Schmidt2010} is 3.5--6.0\,mag, broader and slightly
redder than our model ranges (the dashes lines in our Fig.5). We note that the
Hammer spectral--typing procedure employed by \citet{Schmidt2010},
was developed by \citet{Covey2007}, who sought to optimize only for K and M dwarfs.
For L dwarfs, the uncertainties should be $\sim$2 subclasses or larger. In
addition, the Hammer is designed for solar-metallicity dwarfs, while Praesepe
has a metallicity of 0.27$\pm$0.10\,dex\,\citep{Pace2008}, hence will be
prone to significant, systematic errors\,\citep{Covey2007}. We therefore keep the
use of our model colours alone for candidate selection for a homogeneous study.
We found that if we use the observed color range from
\citet{Schmidt2010} instead, the observed rise in the mass function around
60\,$M_{\rm Jup}$ (cf. Section 4) remains -- our main conclusion remains unchanged.

\begin{figure}
  \centering
  \includegraphics[width=8cm]{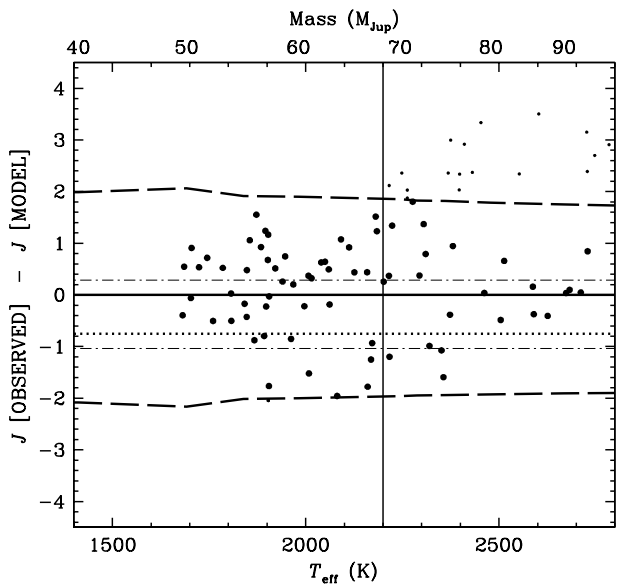}
  \caption{\label{fig:mj_vs_mj} Difference between the observed $J$
    magnitude and the model $J$ magnitude computed from the derived
    mass and $T_{\rm eff}$, as a function of $T_{\rm eff}$. The
    vertical line marks the location of L0 dwarfs. The dotted line
    (at $-0.753$\,mag) represents the offset due to the possible
    presence of unresolved binaries, the dashed-dotted lines
    represent the error in the magnitude determination, and the
    long-dashed lines represent the uncertainties in the age
    and distance of Praesepe. The horizontal solid line just traces
    zero.}
\end{figure}

\begin{figure}
  \centering
  \includegraphics[width=8cm]{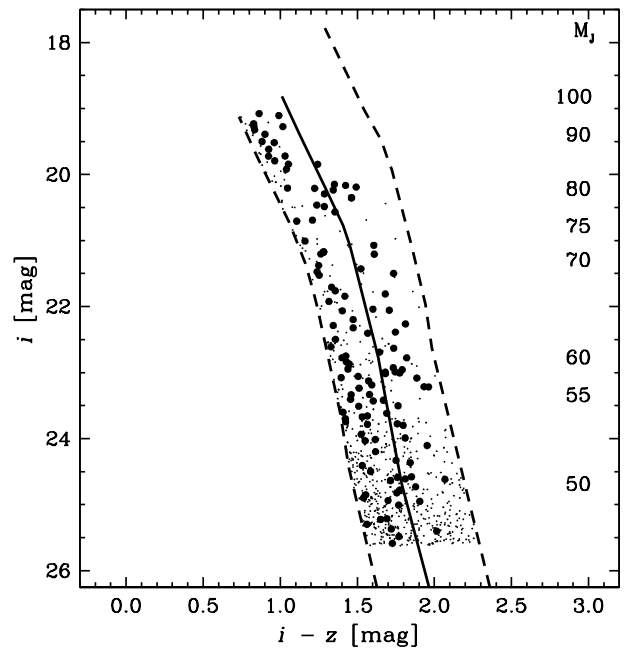}
  \caption{\label{fig:cmd-iz_izJK}Same as Fig.~\ref{fig:cmd-iz}, but with 
    large dots standing for the 59 final 
    candidates, and small dots for the 709 candidates that pass the first 
    selection procedures. 
    }
\end{figure}

\begin{table*}
\caption{The 22 low mass and 37 brown dwarf candidates in Praesepe.}
\label{tab02}
\begin{center}
\begin{tabular}[longtable]{lllccccccccccc}
\hline
\noalign{\smallskip}
ID &RA(J2000)&DEC(J2000)&$i $ &err&$z $ &err&$J$&err&$K_{\rm s}$&err&$T_{\rm eff}$&  $M$ & $J_{\rm model}$\\
      &         &          &(mag)     &   &(mag)     &   &(mag)& &(mag)      &   & (K)&($M_{\rm Jup})$& (mag) \\
\noalign{\smallskip}
\hline
\noalign{\smallskip}
A01 & 8:40:12.599 & 19:56:50.33 & 20.29 & 0.03 &  19.01 & 0.06 & 18.14 & 0.05 &  17.01 & 0.05 & 3023 & 109 &  16.39  \\
A02 & 8:40:18.893 & 19:57:07.40 & 20.17 & 0.02 &  18.74 & 0.06 & 17.75 & 0.03 &  16.77 & 0.04 & 2729 &  93 &  16.90  \\
A03 & 8:40:20.624 & 19:43:40.72 & 23.42 & 0.06 &  21.75 & 0.10 & 19.55 & 0.11 &  17.88 & 0.10 & 1921 &  58 &  19.04  \\
A04 & 8:40:24.901 & 19:57:15.52 & 22.93 & 0.05 &  21.19 & 0.03 & 19.25 & 0.18 &  17.53 & 0.08 & 1940 &  59 &  18.99  \\
A05 & 8:40:25.044 & 19:41:44.56 & 19.23 & 0.02 &  18.40 & 0.02 & 17.37 & 0.02 &  16.47 & 0.03 & 3093 & 113 &  16.27  \\
A06 & 8:40:25.773 & 19:57:08.28 & 19.84 & 0.03 &  18.79 & 0.04 & 17.63 & 0.04 &  16.72 & 0.04 & 2821 &  98 &  16.74  \\
A07 & 8:40:25.981 & 19:57:07.02 & 19.72 & 0.03 &  18.69 & 0.04 & 17.55 & 0.04 &  16.63 & 0.04 & 2900 & 102 &  16.61  \\
A08 & 8:40:27.459 & 19:43:19.69 & 22.06 & 0.08 &  20.66 & 0.08 & 18.60 & 0.04 &  16.96 & 0.04 & 1997 &  61 &  18.82  \\
A09 & 8:40:32.282 & 19:43:53.72 & 25.33 & 0.15 &  23.23 & 0.07 & 20.28 & 0.20 &  18.84 & 0.24 & 1855 &  56 &  19.22  \\
A10 & 8:40:33.639 & 19:53:08.76 & 23.94 & 0.03 &  22.41 & 0.02 & 19.77 & 0.06 &  18.37 & 0.08 & 1902 &  57 &  19.09  \\
A11 & 8:40:35.313 & 19:44:54.81 & 19.72 & 0.01 &  18.80 & 0.02 & 16.66 & 0.01 &  15.83 & 0.02 & 2351 &  75 &  17.73  \\
A12 & 8:40:44.975 & 20:00:24.80 & 23.43 & 0.02 &  21.83 & 0.02 & 19.09 & 0.02 &  17.56 & 0.04 & 1842 &  55 &  19.26  \\
A13 & 8:40:49.194 & 19:59:12.88 & 23.19 & 0.02 &  21.59 & 0.03 & 19.31 & 0.04 &  18.41 & 0.08 & 2040 &  62 &  18.68  \\
A14 & 8:41:08.239 & 19:47:08.93 & 23.13 & 0.09 &  21.55 & 0.04 & 19.05 & 0.02 &  17.69 & 0.03 & 1905 &  57 &  19.08  \\
A15 & 8:41:15.526 & 19:44:11.43 & 20.49 & 0.02 &  19.20 & 0.01 & 17.00 & 0.07 &  16.13 & 0.03 & 2169 &  67 &  18.25  \\
A16 & 8:41:24.916 & 19:57:26.19 & 19.11 & 0.01 &  18.12 & 0.06 & 16.13 & 0.01 &  15.25 & 0.01 & 2356 &  75 &  17.72  \\
A17 & 8:41:26.483 & 19:51:59.73 & 21.38 & 0.01 &  20.13 & 0.02 & 18.24 & 0.02 &  17.34 & 0.02 & 2294 &  72 &  17.86  \\
B01 & 8:38:39.176 & 19:42:54.85 & 20.21 & 0.01 &  18.99 & 0.01 & 17.30 & 0.03 &  16.26 & 0.02 & 2373 &  76 &  17.68  \\
B02 & 8:38:59.208 & 20:02:32.63 & 22.99 & 0.03 &  21.31 & 0.01 & 19.29 & 0.12 &  17.95 & 0.10 & 2050 &  63 &  18.65  \\
B03 & 8:39:06.931 & 19:51:00.19 & 19.28 & 0.01 &  18.26 & 0.09 & 16.98 & 0.01 &  16.13 & 0.02 & 2712 &  92 &  16.93  \\
B04 & 8:39:28.285 & 19:44:01.63 & 19.62 & 0.01 &  18.69 & 0.09 & 17.64 & 0.04 &  16.66 & 0.03 & 3093 & 113 &  16.27  \\
B05 & 8:39:34.775 & 19:43:06.65 & 22.61 & 0.05 &  21.28 & 0.03 & 19.11 & 0.07 &  17.77 & 0.09 & 2060 &  63 &  18.61  \\
B06 & 8:39:37.010 & 19:52:30.06 & 22.20 & 0.02 &  20.72 & 0.03 & 18.32 & 0.05 &  16.68 & 0.03 & 1892 &  57 &  19.12  \\
B07 & 8:39:38.172 & 20:01:17.52 & 19.32 & 0.01 &  18.49 & 0.01 & 17.04 & 0.02 &  16.06 & 0.02 & 2674 &  90 &  17.00  \\
B08 & 8:39:51.841 & 19:50:42.94 & 22.63 & 0.02 &  20.90 & 0.03 & 18.31 & 0.05 &  16.84 & 0.03 & 1867 &  56 &  19.19  \\
B09 & 8:39:56.429 & 19:47:31.41 & 22.96 & 0.03 &  21.17 & 0.04 & 19.36 & 0.09 &  17.95 & 0.11 & 2112 &  65 &  18.44  \\
C01 & 8:38:34.501 & 19:31:08.83 & 19.27 & 0.01 &  18.44 & 0.01 & 16.81 & 0.09 &  16.13 & 0.02 & 2590 &  86 &  17.18  \\
C02 & 8:38:48.106 & 19:31:43.17 & 23.21 & 0.06 &  21.28 & 0.02 & 19.73 & 0.10 &  18.14 & 0.12 & 2181 &  68 &  18.21  \\
C03 & 8:38:49.426 & 19:16:43.34 & 23.50 & 0.03 &  21.74 & 0.02 & 18.98 & 0.06 &  17.34 & 0.05 & 1808 &  54 &  19.49  \\
C04 & 8:38:59.835 & 19:31:52.00 & 21.71 & 0.01 &  20.37 & 0.01 & 18.62 & 0.04 &  17.59 & 0.07 & 2310 &  73 &  17.83  \\
C05 & 8:39:00.756 & 19:37:38.70 & 19.08 & 0.01 &  18.21 & 0.01 & 16.70 & 0.01 &  15.77 & 0.01 & 2626 &  88 &  17.10  \\
C06 & 8:39:02.279 & 19:37:38.47 & 22.32 & 0.01 &  20.85 & 0.01 & 18.72 & 0.07 &  17.88 & 0.08 & 2159 &  67 &  18.29  \\
C07 & 8:39:11.773 & 19:36:35.63 & 21.53 & 0.01 &  20.27 & 0.01 & 18.61 & 0.10 &  17.44 & 0.06 & 2381 &  76 &  17.66  \\
C08 & 8:39:22.309 & 19:38:37.41 & 22.06 & 0.01 &  20.35 & 0.01 & 18.42 & 0.10 &  16.96 & 0.04 & 2062 &  63 &  18.61  \\
C09 & 8:39:35.180 & 19:34:13.22 & 19.97 & 0.01 &  18.94 & 0.01 & 16.50 & 0.10 &  16.41 & 0.03 & 2160 &  67 &  18.28  \\
C10 & 8:39:35.196 & 19:33:50.49 & 20.21 & 0.01 &  19.16 & 0.01 & 17.51 & 0.01 &  16.47 & 0.02 & 2462 &  79 &  17.47  \\
C11 & 8:39:36.737 & 19:34:55.25 & 24.59 & 0.12 &  22.83 & 0.03 & 20.16 & 0.18 &  18.43 & 0.13 & 1786 &  53 &  19.64  \\
C12 & 8:39:47.778 & 19:28:03.10 & 20.15 & 0.01 &  18.80 & 0.01 & 16.90 & 0.01 &  15.82 & 0.01 & 2217 &  69 &  18.09  \\
C13 & 8:39:54.504 & 19:20:09.00 & 23.06 & 0.03 &  21.55 & 0.01 & 19.16 & 0.07 &  18.16 & 0.11 & 2007 &  61 &  18.79  \\
D01 & 8:40:10.438 & 19:24:07.42 & 19.39 & 0.01 &  18.49 & 0.02 & 17.08 & 0.01 &  16.25 & 0.02 & 2683 &  91 &  16.98  \\
D02 & 8:40:13.215 & 19:27:00.51 & 24.64 & 0.07 &  22.92 & 0.04 & 20.14 & 0.19 &  18.06 & 0.11 & 1703 &  50 &  20.21  \\
D03 & 8:40:13.829 & 19:26:55.82 & 24.33 & 0.05 &  22.58 & 0.05 & 20.35 & 0.24 &  18.65 & 0.20 & 1896 &  57 &  19.11  \\
D04 & 8:40:20.994 & 19:38:46.21 & 24.11 & 0.14 &  22.15 & 0.07 & 19.72 & 0.17 &  18.10 & 0.12 & 1848 &  55 &  19.24  \\
D05 & 8:40:21.561 & 19:38:51.23 & 23.88 & 0.15 &  21.81 & 0.06 & 19.31 & 0.12 &  17.40 & 0.06 & 1760 &  52 &  19.81  \\
D06 & 8:40:25.698 & 19:36:25.68 & 19.52 & 0.01 &  18.56 & 0.04 & 16.89 & 0.01 &  16.10 & 0.02 & 2504 &  81 &  17.37  \\
D07 & 8:40:26.220 & 19:37:54.58 & 23.70 & 0.08 &  22.28 & 0.05 & 19.71 & 0.13 &  18.44 & 0.16 & 1947 &  59 &  18.97  \\
D08 & 8:40:39.292 & 19:28:39.49 & 22.04 & 0.02 &  20.44 & 0.01 & 18.07 & 0.02 &  16.76 & 0.02 & 1962 &  59 &  18.93  \\
D09 & 8:40:43.705 & 19:29:52.89 & 25.40 & 0.17 &  23.39 & 0.10 & 21.10 & 0.38 &  18.66 & 0.14 & 1705 &  50 &  20.19  \\
D10 & 8:40:50.371 & 19:29:46.29 & 22.55 & 0.05 &  21.22 & 0.04 & 18.88 & 0.05 &  17.20 & 0.04 & 1898 &  57 &  19.11  \\
D11 & 8:40:53.452 & 19:19:13.98 & 23.02 & 0.03 &  21.33 & 0.03 & 18.82 & 0.04 &  17.10 & 0.03 & 1847 &  55 &  19.25  \\ 
D12 & 8:40:53.559 & 19:41:00.05 & 19.85 & 0.02 &  18.60 & 0.03 & 16.82 & 0.05 &  15.61 & 0.02 & 2320 &  73 &  17.81  \\
D13 & 8:40:56.008 & 19:25:31.85 & 24.62 & 0.10 &  22.55 & 0.06 & 20.01 & 0.12 &  17.67 & 0.06 & 1682 &  50 &  20.39  \\
D14 & 8:41:00.974 & 19:32:00.88 & 23.01 & 0.08 &  21.23 & 0.06 & 19.08 & 0.06 &  17.81 & 0.07 & 2015 &  61 &  18.76  \\
D15 & 8:41:02.632 & 19:22:13.66 & 23.66 & 0.04 &  22.09 & 0.04 & 20.22 & 0.16 &  19.35 & 0.25 & 2216 &  69 &  18.10  \\
D16 & 8:41:06.062 & 19:27:47.71 & 24.41 & 0.08 &  22.88 & 0.06 & 20.26 & 0.17 &  19.41 & 0.30 & 1903 &  57 &  19.09  \\
D18 & 8:41:15.768 & 19:28:15.81 & 19.79 & 0.01 &  18.83 & 0.02 & 17.35 & 0.01 &  16.49 & 0.02 & 2588 &  86 &  17.19  \\
D19 & 8:41:17.045 & 19:28:13.46 & 22.84 & 0.02 &  21.41 & 0.02 & 19.44 & 0.08 &  18.38 & 0.12 & 2184 &  68 &  18.20  \\
D20 & 8:41:32.924 & 19:32:13.40 & 24.01 & 0.03 &  22.40 & 0.04 & 19.52 & 0.09 &  17.86 & 0.07 & 1807 &  54 &  19.49  \\
D21 & 8:41:36.404 & 19:25:06.45 & 20.71 & 0.01 &  19.60 & 0.01 & 18.01 & 0.03 &  17.24 & 0.05 & 2513 &  82 &  17.35  \\
\noalign{\smallskip}
\hline
\noalign{\smallskip}
\end{tabular}
\end{center}
\end{table*}

\section{\label{results-survey} Results}

\subsection{Selected photometric candidates}

We find that 59 photometric candidates survive the selection procedures (based on isochrones
assuming dusty atmospheres), a density of about 100 objects per square
degree. Details of all photometric candidates are listed in Table~3. The
identification number (ID) of a candidate is defined according to the field in
which it was found and a sequential number for that field. The last column,
$J_{\rm model}$, is the predicted $J$ magnitude based on photometric
determination of $T_{\rm eff}$ and mass.

Our survey concentrates on the substellar regime. Saturation occurs at
$\sim$18~mag in $z$ band, corresponding to $\sim$100\,$M_{\rm Jup}$.
Therefore, most of the low mass candidates discovered in previous surveys
(e.g.\ \citealt{pinfield97}, \citealt{hambly1995}) are saturated in our LBT images.
Only a few faint brown dwarfs classified by \cite{pinfield97},
\cite{gonzalez-garcia2006}, and \cite{boudreault2010} are rediscovered in the
current survey (cf.\ Table~4). These objects are M5--9 dwarfs, according to the
photometric relations given by \citet{West2008}.

Some of our targets were previously identified as cluster members
but rejected by our selection procedures or visual inspection. For example,
eleven of the 150 Boudreault et al.\ (2010) candidates are detected in our
LBT survey (the rest are mostly saturated). However, nine of them are rejected 
in this work (cf.\ Table~5). Among them, seven are rejected on the basis of  
the $z$ vs. $i$--$z$ CMD selection, because they are bluer than the isochrone area.
Another one is obviously not a point-like source in the LBT image, and another
is rejected because its observed $J$ magnitude is inconsistent with its
model-predicted magnitude. The remaining two targets are confirmed to be
cluster dwarf stars. As our current work employed more photometric bands than
\cite{boudreault2010}, it is unsurprising that our selection is more
conservative.

Most of our candidates are in the substellar regime, and other than the
five targets listed in \ Table~4, no other accurate photometric observations
are available from past epochs. This precludes using proper motions as a means
of cluster membership assessment at this time.

\begin{table}
\caption{Previously discovered low-mass Praesepe candidates found in our survey.}
\label{tab04}
\begin{center}
\begin{tabular}[longtable]{llllc}
\hline
\noalign{\smallskip}
ID    &RA(J2000)&DEC(J2000)& Alternative name & Ref.$^{a}$ \\
      &($^{\rm h  m  s}$)&($^{\circ}$ \arcmin \arcsec)&&\\
\noalign{\smallskip}
\hline
\noalign{\smallskip}
A16  & 8:41:24.916 & 19:57:26.19 & NGC2632 PHJ20    & [1]\\
A17  & 8:41:26.483 & 19:51:59.73 & J084126.5+195200 & [2]\\
C12  & 8:39:47.778 & 19:28:03.10 & Praesepe017      & [3]\\
     &             &             & NGC2632 PHJ11    & [1]\\
D12  & 8:40:53.559 & 19:41:00.05 & Praesepe001      & [3]\\
D08  & 8:40:39.292 & 19:28:39.49 & J084039.3+192840 & [2]\\
\noalign{\smallskip}
\hline
\noalign{\smallskip}
\end{tabular}
\begin{list}{}{}
Refs: [1]~\citet{pinfield97}; [2]~\citealt{gonzalez-garcia2006}; 
[3]~\citet{boudreault2010}
\end{list}
\end{center}
\end{table}

\subsection{\label{get-mass} Photometrically-derived masses and effective
temperatures} 

From the evolutionary tracks and atmosphere models described previously, we
obtained the magnitudes $m_{A}$ from the average flux of a star in a specific
band $A$ using the equation 

\begin{equation} {m_{A}=-2.5\;\mathrm{log}\;f_{A}\;+\;c_{A}},
\label{eqw:phot-cal}
\end{equation}

\noindent where $m_{A}$ is the magnitude observed in a given passband,
$f_{A}$ the flux received on Earth in this passband, and
$c_{A}$ is a constant that remains to be determined. The flux
$f_{A}$ was obtained using the total transmission function of the
passband for a given filter, convolved with the quantum efficiency of
the CCDs (we assumed that the telescope and instrumental throughput is
flat over each passband). To transform our optical $i$ and
$z$ band magnitudes from the models to the Johnson photometric
system, we assumed all magnitudes $m_{A}$ to be equal to 0.03 when
$f_{A}$ is the average Vega flux received on Earth. The constant
$c_{A}$ for each passband is then determined using the Vega flux from
\cite{colina1996}. The fitted values for these constants are
$i=-22.180$ and $z=-22.706$\,mag. The values computed by
\cite{boudreault2010} for $J$ and $K_{\it s}$ were $-23.687$ and
$-25.908$\,mag, respectively.

The masses and effective temperatures were estimated in the way
described by \cite{boudreault2009}\footnote{We first normalized the energy
distribution of each object to the energy distribution of the model using the
$J$ filter. The energy distribution was then fitted via a least squares fit of
the model magnitudes to the measured ones.}. For the faintest objects where
only $i$ and $z$ are available, the colour $i$-$z$ is used to compute masses
and $T_{\rm eff}$. There are different sources of errors for the estimation of
the mass and $T_{\rm eff}$, including the photon noise, the photometric
calibration, the least squares fitting (imperfect model), and the uncertainties
in the age of and distance to Praesepe. The latter two are the most significant
errors and give the uncertainties of 0.008\,M$_\odot$ and 263\,K for a
0.05\,M$_\odot$ substellar object (~$T_{\rm eff}=1\,690$\,K~), 0.010\,M$_\odot$
and 260\,K for a 0.06\,M$_\odot$ substellar object (~$T_{\rm eff}=1\,990$\,K~),
and 0.008\,M$_\odot$ and 201\,K for a 0.072\,M$_\odot$ object at the hydrogen
burning limit (~$T_{\rm eff}=2\,293$\,K~).

\begin{table*}
\caption{The 9 Praesepe cluster candidates identified by \citet{boudreault2010} that were rejected in this work 
by colour or brightness. The IDs in the first column are taken from \citet{boudreault2010}. The fourth object 
is not a point-like source in the LBT image, thus no optical photometry measurement is available. }
\label{tab05}
\begin{center}
\begin{tabular}[]{lllccccccccccc}
\hline
\noalign{\smallskip}
ID   &RA(J2000)    &DEC(J2000)   &$i $  & $z $  & $I_c$  & $J$   &$K_{\rm s}$ & $M$             & $T_{\rm eff}$ & $J_{\rm model}$\\
     &             &             &(mag) & (mag) & (mag)  & (mag) & (mag)      & ($M_\odot)$ & (K)           & (mag) \\
\noalign{\smallskip}
\hline
\noalign{\smallskip}
005 & 8:41:08.50  & +19:54:02.0  & 20.19  & 18.70 &  19.02  & 16.58 &  15.39  & 0.088  & 2636 &  17.06  \\
007 & 8:39:39.56  & +19:47:54.3  & 18.93  & 17.80 &  17.95  & 16.10 &  15.07  & 0.104  & 2860 &  16.58  \\
009 & 8:39:55.84  & +19:53:14.3  & 18.05  & 17.78 &  20.29  & 17.50 &  16.54  & 0.081  & 2520 &  17.32  \\
018 & 8:39:42.79  & +19:35:48.2  & --     & --    &  18.27  & 16.20 &  15.21  & 0.097  & 2782 &  16.78   \\%
022 & 8:41:04.20  & +19:31:27.8  & 20.12  & 19.15 &  18.89  & 16.67 &  15.75  & 0.092  & 2702 &  16.97  \\
901 & 8:39:59.45  & +19:43:37.4  & 18.05  & 18.30 &  19.09  & 17.16 &  16.41  & 0.084  & 2574 &  17.20  \\
902 & 8:39:23.72  & +19:52:01.8  & 20.41  & 19.46 &  20.15  & 17.77 &  16.88  & 0.073  & 2348 &  17.72  \\
903 & 8:40:00.20  & +19:30:27.0  & 19.59  & 19.01 &  19.74  & 17.50 &  16.62  & 0.076  & 2409 &  17.57  \\
914 & 8:38:52.02  & +19:35:05.3  & 19.48  & 18.69 &  19.12  & 17.25 &  16.35  & 0.085  & 2591 &  17.16  \\

\noalign{\smallskip}
\hline
\noalign{\smallskip}
\end{tabular}
\end{center}
\end{table*}

\subsection{Contamination by non-members}

As mentioned above, the three main sources of contamination are
background red giants, unresolved galaxies, and Galactic M and L
dwarfs. Red giants contaminate the high mass end of this study, as seen in the
$i-J$ vs. $z-K_{\it s}$ diagram, hence can be ignored. Although some
types of galaxies have similar colours to Praesepe cluster members more
massive than 60\,$M_{\rm Jup}$, these low-redshift galaxies are in general
extended sources and therefore easily rejected by our visual inspection. Among
the 74 candidates that passed our selection procedures, we identify four as
galaxies on the basis of their LBT images. Other possible sources are field L dwarfs
and high redshift quasars (for instance at redshift $z\sim$6;
\citealt{caballero2008}). However, because these quasars have spectral energy
distributions similar to mid-T dwarfs, whereas our faintest candidates have
colours of early L dwarfs, and given that they are rare (0.25 quasars at
$5.5<z<6.5$ in a 0.59\,deg$^{2}$ survey, Stern et al. 2007), the MF should not
be affected by quasar contamination. 

The contamination by field dwarfs is not negligible.  \citet{caballero2008} 
identified possible field dwarf contaminants covering spectral types from M3 to
T8 from the literature and presented the spatial density in the solar
neighbourhood in their Table~3. From this, the spatial density of field dwarfs
in the vicinity of Praesepe can be easily inferred, given the Galactic latitude
of Praesepe of $b=+32.5$\,deg and its distance of 190\,pc, assuming an
exponential decrease for stellar density perpendicular to the Galactic disk
with a scale height of 500\,pc. The absolute $J$ band magnitude range
constrained by our selection procedures is $\sim\pm 2$~mag, as shown in
Fig.~\ref{fig:mj_vs_mj}, corresponding to a certain distance interval and a
survey `volume', which is defined by the product of survey area and depth. We
calculated the number of contaminants by multiplying the survey volume by spatial
density at each mass bin. The result is shown in Table~\ref{dwarf_contamintes}.
The first column gives the central value of log\,$M$ in each interval,
while the second and third columns present the corresponding $T_{\rm eff}$ and
$J_{\rm model}$ values at that mass. The fourth column is the number density of
cluster member candidates (also shown Fig.~\ref{fig:mf-prae-us} as filled
triangles) after applying all corrections (except for in the lowest two bins,
where contaminations are too high to be corrected). The final two columns give
the number density of field dwarf contaminants and corresponding fraction. }
We found that the field dwarf contaminants do not affect the MF shape. The
contamination is significant for $J_{\rm model}\ga 20$~mag. At this magnitude,
the \citet{boudreault2010} $J$ band has a completeness of 88\%. Therefore,
below this magnitude, the mass functions we derived are probably upper limits.

\begin{table}
\caption{Number density of the cluster member candidates.}
\label{dwarf_contamintes}
\begin{center}
\begin{tabular}{cccccc}
\hline
\hline
log\,$M$       & $T_{\rm eff}$& $J_{\rm model}$    & N(cand.) & N(cont.)& Fraction\\ 
($M_{\rm Jup}$)& (K)          &(mag)   & (deg$^{-2}$) &(deg$^{-2}$)      \\
\noalign{\smallskip}
\hline
\noalign{\smallskip}
 1.625       & 1412        & 23.29 &  13  & 195   &  1500\%\\
 1.675       & 1692        & 20.28 &  27  &  41   &  150\%\\
 1.725       & 1839        & 19.27 & 18   &   3.3 &  15\%\\
 1.775       & 1981        & 18.87 & 33   &   4.8 &  13\%\\
 1.825       & 2244        & 18.01 & 38   &   2.3 &  5.7\%\\
 1.875       & 2361        & 17.71 & 17   &   3.2 &  16\%\\
 1.925       & 2479        & 17.43 & 8.4  &   1.7 &  17\%\\
 1.975       & 2668        & 17.01 & 8.4  &   1.6 &  16\%\\
\noalign{\smallskip}
\hline
\noalign{\smallskip}
\end{tabular}
\end{center}
\end{table}

We conclude that the various contaminants are either negligible, or do not
affect the MF shape in the range that we can investigate quantitatively, i.e.,
from about 53 to 94\,$M_{\rm Jup}$.

\section{\label{mf} Mass function of very low mass and substellar population of 
Praesepe}

The mass function, $\xi$(log$_{10}$M), we present here is the total number of
objects per square degree in each logarithmic mass interval log$_{10}$M to
log$_{10}$M + 0.1. Since we do not make any corrections for binaries, we
compute here a \textit{system} MF.

As our candidates have been selected only from their photometric properties,
cluster membership confirmation via spectroscopy is desirable. However, these  
observations will not be feasible in the near future because of the faintness of
our candidates. The following discussion is therefore based on the assumption
that the MF of candidates is similar to that of `real' cluster members. The
assumption is possibly valid because our derived MF is consistent with that
given by Boudreault et al. (2010) in the common mass bin, and the contamination
by field dwarfs, giants, and galaxies should not affect the shape of the MF
significantly.

To account for the survey detection efficiency, we use a simple simulation. For
example, to calculate the detection efficiency of candidate A01 in $i$ band, we
select a bright but unsaturated star in the $i$ band CCD image in which A01
resides, scale it down to the magnitude of A01 (i.e., $i=20.29$), and
randomly cast this `fake' star in the CCD image 100 times. We then search and
re-measure the `fake' star with our procedures. The detection efficiency is the
fraction of `fake' stars that have been re-discovered. We run this test for
each filter ($izJK_{\it s}$) and multiply the detection efficiencies together
(as we need a detection in every filter) to evaluate the overall detection
efficiency for each candidate. This detection efficiency (or recovery rate) is
about 90\% for the brightest candidates and drops very quickly to 10\% for
the faintest candidates. 

\begin{figure}
\centering
\includegraphics[width=8cm]{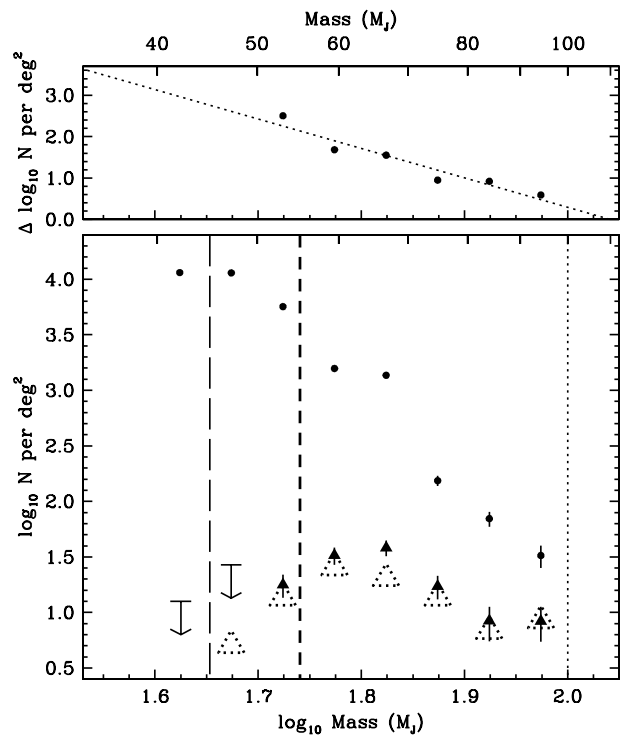}
\caption{\label{fig:mf-prae-us} \textit{Lower panel.} Mass function based on (a)
the LBT $i$ and $z$ photometry (filled dots) after detection efficiency 
corrections, and (b) the LBT photometry plus the
$J$ and $K_s$ photometry from \cite{boudreault2010} (dotted open triangles: the
original MF; filled triangles: the MF after detection efficiency corrections 
and removing contaminants). The
selection and mass calibration are based on dusty atmospheres. Error bars are
Poissonian arising from the number of objects in each bin, except for the lowest-mass 
bin, for which the error bar is from the linear fit in the top panel. The vertical
thin dotted lines are the mass limits at which detector saturation occurs in
the $i$ and $z$-bands. The vertical thin long dashed line is the mass at the
5$\sigma$ detection limit of our optical LBT data while the thick short dashed
line is the mass at the 5$\sigma$ detection limit of the NIR data of
\cite{boudreault2010}. \textit{Top panel.} Difference of the log of the number of objects
(in each mass bin) between the MF computed using the optical $iz$ data
and the MF computed using both $iz$ and 
NIR $JK_{\it s}$ data from \cite{boudreault2010}. The dotted line is a
linear fit.}
\end{figure}

We mentioned in Section \ref{obs-data-calib} that our optical photometry
reaches lower masses than the NIR photometry that we used. To compute the MF of
Praesepe to the lowest mass bin reached without optical data, we first computed
a MF\footnote{Note that this is an inaccurate `MF', because of serious
contaminations.} using only the optical $iz$ photometry. This `MF' is presented
in Fig.~\ref{fig:mf-prae-us} as filled dots. We computed a second MF from the
list of candidates that pass the three selection criteria and are also
detected in the survey of \cite{boudreault2010} in the NIR $J$ and $K_{\it s}$
bands (presented on Fig.~\ref{fig:mf-prae-us} as filled triangles). For each
mass bin, we computed the number of objects removed as a result of adding the $J$
and $K_{\it s}$ filters to our selection process and mass determination procedure
(plotted as a function of mass in Fig.~\ref{fig:mf-prae-us}, top panel). We
fitted a linear function to estimate the number of objects that would be
removed \textit{if} we had additional $J$ and $K_{\it s}$ photometry to
40--45\,$M_{\rm Jup}$, which is our lowest mass data point in
Fig.~\ref{fig:mf-prae-us}. However, as shown in Table~6, the contamination in
the two lowest mass bins is so overwhelming that the MFs in these two bins can
only be regarded as upper limits and are no longer discussed in the
paper. 

Our derived MF (presented in Fig. 8, 9, \& 10) shows a rise from
105\,$M_{\rm Jup}$ to 67\,$M_{\rm Jup}$ and then a turn-over at
$\sim$67\,$M_{\rm J}$. This turn-over occurs well above the 5$\sigma$ of either
$iz$ bands or $JK_{\it s}$ bands (e.g.\ at $\sim$67\,$M_{\rm Jup}$, $i\sim$22)
and we note that we have corrected the incompleteness of our survey and field dwarf
contaminations. We therefore believe that this feature is genuine. This is the
first time a clear rise in the substellar MF in an old open cluster has been observed.

\begin{figure} \centering
  \includegraphics[width=8cm]{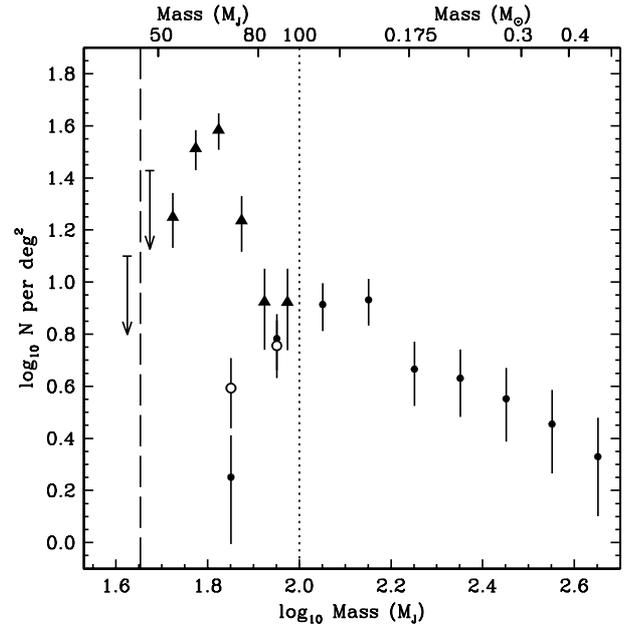}
  \caption{\label{fig:mf-praesepe} MF of Praesepe based on our
    survey LBT $iz$ and $\Omega$2k $JK_s$ photometry (triangles), compared with 
    that from \cite{boudreault2010} (\textit{open dots}
    assuming a dusty atmosphere and \textit{filled dots} assuming a
    dust-free atmosphere). Error bars are Poissonian arising from the
    finite number of objects observed in each bin, except for the last bin, 
    for which the error bar is derived from the linear fit. The vertical thin dotted
    line is the mass limit above which detector saturation occurs in
    the $i$ and $z$--bands. The vertical thin long-dashed line is the
    mass at the 5$\sigma$ detection limit of our optical LBT data.}
\end{figure}

The MF of Praesepe near the hydrogen-burning limit was previously obtained
in several studies. However, only \cite{boudreault2010} provide a
common mass range for comparison, as shown in Fig.~\ref{fig:mf-praesepe}. 
In the first substellar mass bin ($\sim$80\,$M_{\rm
Jup}$) we see that, both surveys give consistent results within their error bars. However, for the
second bin at $\sim$70\,$M_{\rm Jup}$, our MF is much higher than that of 
\cite{boudreault2010}; the discrepancy is smaller when considering the MF from 
that work using the dusty 
atmosphere (open dots), which is still a reasonable model for
such low mass stars ($\sim$M9/L0). This may indicate that some faint candidates
are missing in the \cite{boudreault2010} survey, as these authors did not make any corrections
for the detection efficiencies. 

We emphasize that our LBT survey covers the very central 0.59\,deg$^2$ of
Praesepe, while the $\Omega$2k survey by \cite{boudreault2010} covers a much
wider area ($\sim$3.1\,deg$^2$). If no significant candidates are missing in
the $\Omega$2k survey, this discrepancy may suggest that the very low mass
cluster members are mostly concentrated in the cluster centre, in contrast to what
is expected from a `dynamical evaporation' of brown dwarf in open clusters.
The basic idea of dynamical evaporation is that lower mass stars in a cluster
have higher speeds according to equipartition of energy, so are able to move
higher in the gravitational potential well of the cluster. Hence the fraction
of low mass stars should increase with increasing distance from the cluster
centre. By comparing the Praesepe and Hyades MFs, \cite{boudreault2010}
concluded that Praesepe might have been less affected by dynamical evolution.

\begin{figure} \centering
  \includegraphics[width=8cm]{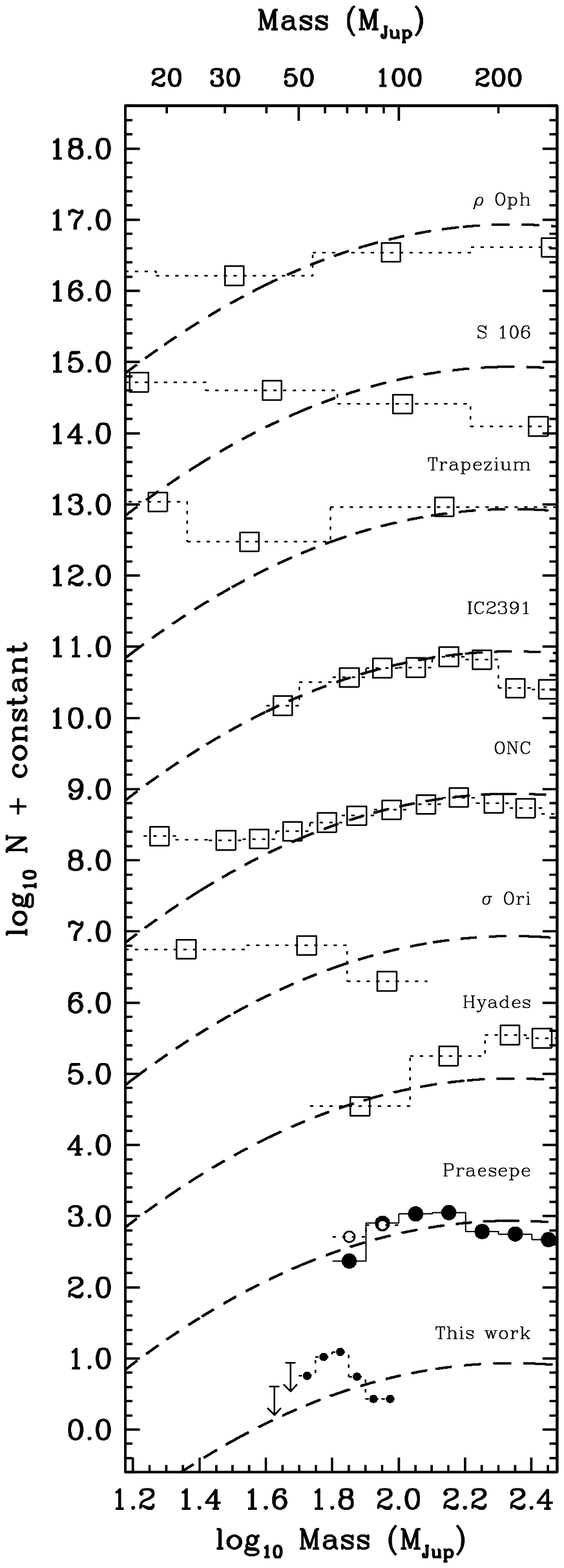}
  \caption{\label{fig:all-mf} Mass functions of various open clusters.
    From top to bottom: $\rho$\,Oph ($\sim$1\,Myr, \citealt{Marsh2010});
    S\,106 ($\sim1$\,Myr, \citealt{oasa2006}, their Fig.~10, middle panel for example);
   Trapezium ($\sim$0.8\,Myr, \citealt{muench2002});
    IC2391 ($\sim$50\,Myr, \citealt{boudreault2009});
    ONC ($\sim$5\,Myr, \citealt{hillenbrand2000});
    $\sigma$ Ori ($\sim$3\,Myr, \citealt{bihain2009});
    Hyades ($\sim$625\,Myr, \citealt{bouvier2008});
    Praesepe ($\sim$590\,Myr, \citealt{boudreault2010} and from this work). 
    We also show the lognormal fit to the Galactic field star MF from \cite{chabrier2003} as 
    a dashed line. All the other MFs are normalized to this at the substellar limit ($\sim$72\,$M_{\rm Jup}$).
}
\end{figure}

Owing to its large distance and old age, no other published MF determination of
Praesepe has reached masses below 70\,$M_{\rm Jup}$. We therefore compare our
results with those from other clusters in Fig.~\ref{fig:all-mf}. This includes
IC\,2391 from \cite{boudreault2009}, ONC from \cite{hillenbrand2000}, $\sigma$
Orionis from \citep{caballero2007,bihain2009}, and the Hyades from
\cite{bouvier2008}. The MF of Praesepe is quite different from both IC\,2391
(age of $\sim$50\,Myr) and the Hyades ($\sim$625\,Myr). Either the `dynamical
evaporation' does not have (or has not yet had) the same effect on these three
clusters, or they had different initial mass functions. Another possibility is
that Praesepe has a different binary fraction. Employing different
cluster member selection criteria may also account for the observed MF
discrepancies among clusters. Further studies are necessary to clarify these points.

The continuing rise of the MF into the substellar regime that we observe has
also been observed in young clusters (as shown in Fig.~\ref{fig:all-mf}),
especially in $\sigma$\,Orionis (\citealt{bihain2009}), Trapezium \citep[turn-over
at $\sim$10--20\,$M_{\rm Jup}$, ][]{muench2002}, $\rho$\,Oph \citep[MF rising to
$\sim$10\,$M_{\rm Jup}$, ][]{Marsh2010}, and in the very low luminosity young
cluster in S\,106, where the MF increases or at least remains flat down to
$\sim$10\,$M_{\rm Jup}$~\citep{oasa2006}. If we assumed a universal IMF, then
it seems that the substellar MF of Praesepe has not evolved significantly since
the cluster formed. 

\section{\label{conclusions} Conclusions}

We have carried out the deepest survey to date of the old open cluster
Praesepe, covering the central 0.59\,deg$^2$ in the $rizY$ bands. The survey
probed a mass range from $\sim$100 to 40$M_{\rm Jup}$ at 5$\sigma$ detection
limit, with which we have derived the very low mass and substellar mass
function of this cluster.

We compared our optical $iz$-bands data, combined with the $\Omega$2k
NIR ($J$ and $K_{\rm s}$) band observations from
\citet{boudreault2010}, with theoretical loci of cluster members based
on a dusty atmosphere (the AMES-Dusty model), to select
cluster member candidates. Our final sample comprises 59 photometric
candidates. We estimate that the contamination by field dwarfs is
about 15\%, and that this does not affect the shape of MF. The
contamination by galaxies and red giants is believed to be
negligible. About two thirds of our cluster members have theoretical
masses below the hydrogen-burning limit at 0.072\,$M_{\odot}$, and are
therefore brown dwarf candidates. We emphasize that to claim
cluster memberships for the candidates, follow-up astrometric or
spectroscopic observations are required. However, given that the
candidates are generally faint and these observations are very
time-consuming, none of them has yet been confirmed in this way. The
discussion in this contribution therefore refers to the mass
function of photometric cluster member {\em candidates}.

The mass function we have inferred for the central 0.59\,deg$^2$ of
Praesepe is consistent with that inferred for a wider area by
\cite{boudreault2010} at a mass just below the substellar boundary, but
deviates by $\sim$0.6\,dex in the next lowest mass bin, which may indicate that there is
either a significant number of objects missing in the Boudreault et al.\ 2010
survey, or a higher concentration of substellar objects in the centre of
Praesepe (as the Boudreault et al.\ survey is at a larger cluster radius). The
latter possibility suggests that the dynamical evolution of very low mass stars
is inefficient in this cluster, as proposed by \cite{boudreault2010} for
explaining the discrepancy between the Praesepe MF and Hyades MF.

The steady rise in the Praesepe MF down to $\sim$70\,$M_{\rm Jup}$ and a
turn-over there were unexpected for this old cluster. Such a significant peak has
never been observed in any other cluster older than 50\,Myr, but has been
observed in several very young open clusters such as $\sigma$\,Orionis or
clusters in star--forming regions (e.g., Trapezium). This suggests that the
dynamical interactions in Praesepe have very little effect on MFs, if we assume
there is a universal initial MF.

\begin{acknowledgements}

This project was supported by DFG-Sonderforschungsbereich 881 ``The Milky Way
System''. Some of the observations on which this work is based were obtained
during LBT programme "LBT-F08-02". Some data analysis in this article has made
use of the freely available R statistical package, http://www.r-project.org.
This research has made use of the SIMBAD database, operated at CDS, Strasbourg,
France. This publication makes use of data products from the Two Micron All Sky
Survey, Sloan Digital Sky Survey, and United Kingdom Infrared Telescope
Infrared Deep Sky Survey.

 \end{acknowledgements}

\end{document}